# Abnormal behavior of crystalline methane in temperature interval 60-70 K. From experiment to theory.

A.V. Leont'eva, A.Yu.Prokhorov
Donetsk Physics & Engineering Institute Of NAS of Ukraine, 72 R.Luxemburg str., Donetsk, 83114, Ukraine. E-mail: vesta-news@yandex.ru

A.Yu.Zakharov
Novgorod State University, Velikiy Novgorod, 173003, Russia.
E-mail: anatoly.zakharov@novsu.ru

A.I.Erenburg
Ben-Gurion University of Negev, P.O.B. 653, Beer-Sheva 84105, Israel.
E-mail: erenbura@bgu.ac.il

**Abstract**

The paper presents an analysis of mechanical, structural, thermophysical and spectral properties of solid methane in temperature interval $0.5 \cdot T_{tr} - T_{tr}$ ($T_{tr}$ is the triple point temperature) under equilibrium vapor pressure. It is shown that the anomalies of the studied properties at temperatures 60-70 K have been observed in the body of the reviewed papers. A concept of "topons" (cooperative excitations of rotational degrees of freedom in solid methane) is proposed. A whole of the observed anomalies can be explained in the frames of this concept.

## 1. Introduction.

Solid methane is the lightest representative among the group of simplest molecular crystals, formed by **$CX_4$** tetrahedral molecules with **4-3m** symmetry [1], where **X** are either hydrogen isotopes or halogen atoms or their isotopes.
The development of far cosmos research and discovery of athmospheres at some planets of Solar system have inspired a new large interest to mechanical and thermodynamic properties of solidified gases, in particular, solid metnane [2]. It is nessesary to note that high-temperature range of solid methane is not studied enough. For instance, well-known reviews [3-4] dedicated to solid methane have ignored the repeating anomalies near temperature 65 K.

In this paper, we present a detailed analysis of extensive set of experimental data on mechanical, structural, thermophysical and spectral properties of solid methane in temperature interval $0.5 \cdot T_{tr} - T_{tr}$ ($T_{tr}$ is the triple point temperature) under equilibrium vapor pressure. We show that the body of reviewed papers reveals the presence of anomalies of above properties at temperatures 60-70 K, in the rest the reliable data are absent. On the basis of theoretical treatment of molecule's rotational behavior in solid methane we have demonstrated that a whole of the observed anomalies can be explained by existence of cooperative rotational degrees of freedom in this material.

## 2. Some experimental data on thermodynamic properties of solid methane

Methane is the most examined crystal of **CX$_4$** group. The well-known reviews [3,4] have collected an extensive data on thermodynamic, structural and mechanical characteristics of solid methane. However, the anomalies observed for a series of solid methane properties at temperatures higher than **0.5 T$_{tr}$** have not mentioned in these reviews.
Under equilibrium vapor pressure, crystalline methane (triple point temperature **T$_{tr}$** = 90.67 K [4]) undergoes a $\lambda$-type phase transition at **T$_\lambda$**= 20.48 K [3]. In both phases, low-temperature **II** and high-temperature **I**, carbon atoms form a FCC lattice (space group *Fm3m* [3]). It is commonly supposed that in low-temperature phase hydrogen atoms are ordered in the lattice (expected space group *Fm3c* or *P4-3m* [3]) and can perform certain librational oscillations in relation to a center of inertia of methane molecule, and also reorientational hopping from one equilibrium position to another. In phase **I**, hydrogen atoms entirely loose the ordering and make "hindered spherical rotation" relatively to a gravity center of carbon atom. Volume jump at the phase transition **I-II** is about 0.3 % [3].

## 3. Essential experimental data

A problem of transformation of methane molecule's rotational state in high-temperature phase was firstly established by K.Tomita [5]. In this paper, he studied **NMR** spectra of solid methane and found that (i) half-width of resonance absorption lines **H$_{1/2}$** decreases suddenly at temperatures T> 60 K (see Fig.1), and (ii) time of spin-lattice relaxation $\tau_c$ varies essentially at the same temperatures (see Fig.2). Accounting for this and the fact that activation energy of methane molecule's rotation rises in approximately twenty times in narrow temperature interval, Tomita concluded that methane molecules make an unbraked rotation at temperatures above 60 K.

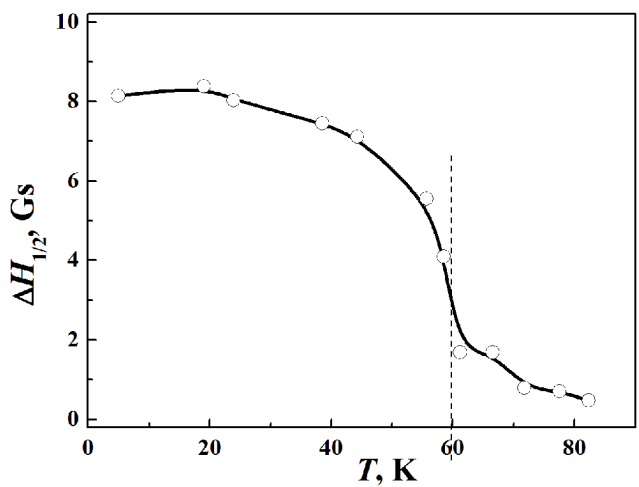

Fig.1. Temperature dependence of half-width of resonance absorption lines $\varDelta H_{1/2}$ of the dynamical local field [5].

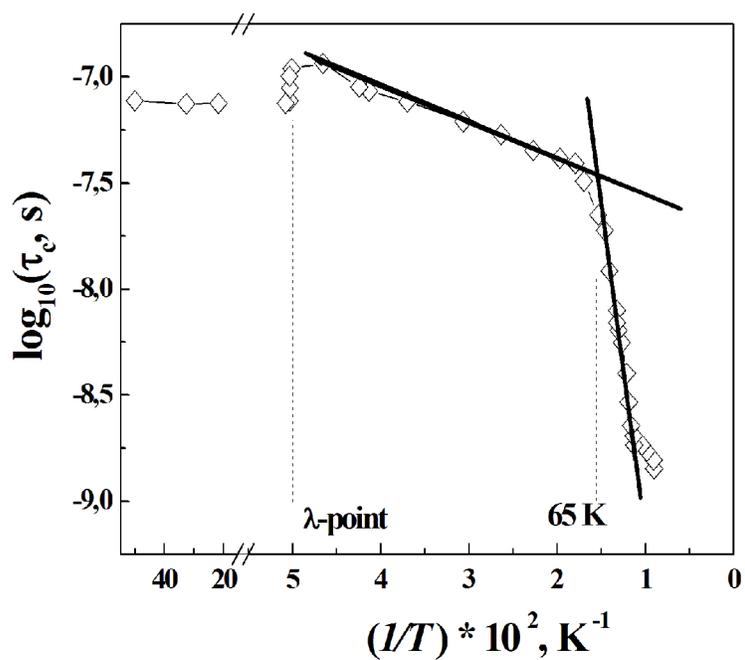

Fig.2. Dependence of characteristic time $\tau_c$ of spin-lattice relaxation on reciprocal temperature [5].

Thus, Tomita experimentally found that rotational state of methane molecules changes suddenly in temperature interval 60-70 K. Such an idea is very non-trivial as it is hard to expect that tetrahedral molecules in solid methane can rotate in general. Nevertheless, a theory of low-temperature diffusion in quantum crystals [6-7] has been created after 16 years after the Tomita's paper. In particular, this theory has revealed that the diffusion coefficient starts to increase with temperature lowering due to quantum effects. Note that this effect was experimentally verified in multiple papers (for instance, [7]). So, from a present-day perspective, an existence of rotational degrees of freedom in the quantum crystal $CH_4$ is not seems absurd. However, intermolecular interactions in solid methane inevitably lead to a collectivization of rotational degrees of freedom. This, in turn, results in renormalization of a sole parameter i.e. moment of inetria I of a spherical top which represents a methane molecule.

Note that such a transformation of rotational behavior of methane molecules displays not only in **NMR** parameters. Fig.3 (a) presents the temperature dependence of low frequency internal friction (**LFIF**) of solid methane $Q^{-1}$ that has been studied by inverse torsion pendulum method at frequencies 5-10 Hz and magnitude of deformation 5 $10^{-5}$. Fig.3 (b) presents the temperature dependence of dynamic shear modulus $G_w$ derived from the frequency of torsion oscillations $f$ ($f^2 \sim G_w$) [8].

It is seen from Fig.3 (a) that a key feature of the **LFIF** spectrum $Q^{-1}(T)$ over a whole temperature range is a large peak at temperatures 65 -70 K. Its magnitude reaches $10^{-1}$, that is by a factor of ten times higher than the magnitude of the peak near temperature of phase I-II transition ($T_\lambda$=20,48К). Such anomalous rise of $Q^{-1}$ in $CH_4$ can not be explained in the frameworks of conventional classic conception. It is necessary to note that such a great **LFIF** peaks were not observed in previous studies of cryocrystals with central intermolecular forces, e.g. in solid argon [9]. This sharp $Q^{-1}$ anomaly is similar to the peaks in **LFIF** spectra of solid oxygen [10]. This allows us to suggest that at T ≈ 65K a latent phase transition in solid methane occurs. It is associated with a sudden transformation of rotational state of $CH_4$ molecules [11]. Later we have shown that the whole set of observed anomalies in $CH_4$ can be successfully explained by existence of cooperative rotational degrees of freedom in this crystal [12].

Fig.3, b shows the temperature dependence of square of torsion oscillations frequency $f^2(T)$ in $CH_4$. It is seen a clearly defined minimum near 65 K indicating a significant alteration of intermolecular forces (in particular, alteration of dynamical shear modulus $G_w$) in narrow temperature interval 3-5 K. Probably, this is related to abovementioned transformation of rotational movement of $CH_4$ molecules in temperature interval 60-70 K which affects the mechanical characteristics of solid methane.

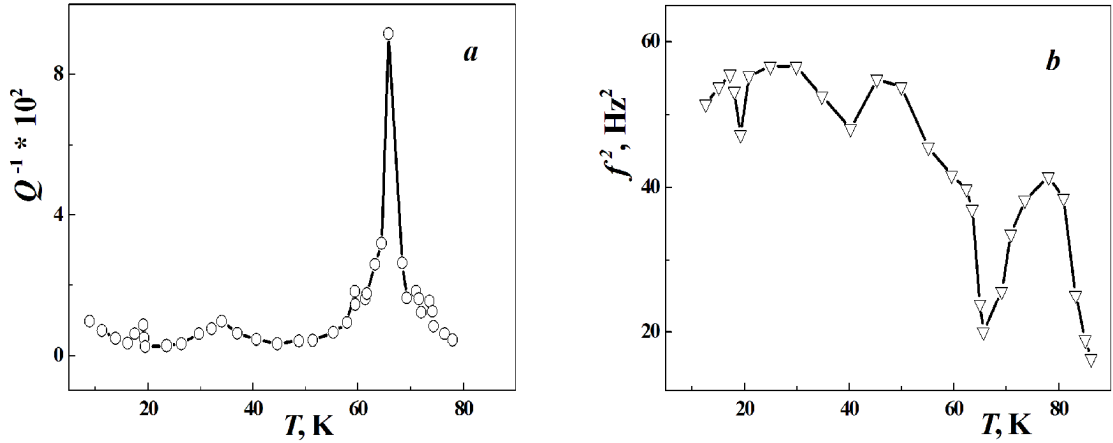

Fig.3. Temperature dependence of **LFIF** $Q^{-1}$ (*a*) and square of torsion oscillations frequency $f^2$ (*b*) of solid methane [8].

Such a conclusion is in agreement with the results of a recent paper [2]. The authors have studied the adhesion and plasticity of polycrystalline methane in temperature interval 10-90 K. They have shown that adhesion is strong in temperature range 50-90 K where solid methane is "soft and sticky". Towards to melting temperature, solid methane behaves as very viscous non-newtonian liquid. At temperatures below 30 K it loses its stickiness and plasticity and behaves as conventional glass.

However, despite of the mentioned temperature interval 50-90 K of softness and plasticity of solid methane, authors of the same paper [2], after studying of temperature dependence of the shear corresponding to a connection breaking point between solid methane and the special probes, have revealed that the shear stress raises with temperature growth up to 60 K, and then rapidly decreases with temperature approaching to melting point (see Fig.6 in [2]). This fact directly indicates the transformation of intermolecular forces character in solid methane that is possibly associated with the transition of cooperative rotational degrees of freedom in solid methane molecules from quantum behavior to classical one [12].

### 4. Analysis of other results and discussion.

Analysis of available experimental data on varios properties of solid methane shows that the transformation of rotational movement of methane molecules is displayed not only in abovementioned **NMR** and mechanical characteristics.
Thus, for example, anomalies of elastic properties of solid methane in temperature interval 60-70 K have been reported in [13,14]. Fig. 4 shows that the anomalies in temperature dependence of longitudinal sound velocity $v_l(T)$ have been observed not only near the phase transition but at temperatures above 60 K as well. Similar peculiarities have been observed in [14] for velocities of both longitudinal $v_l$ and transverse $v_s$ sound (see Fig.5).

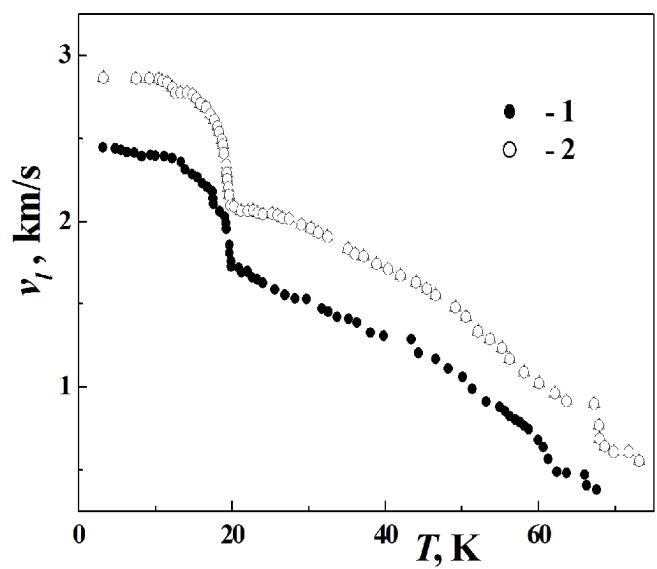

Fig.4. Temperature dependence of longitudional sound velocity $v_l$ [13] for the samples prepared at various growth rates: 1 – fast growth; 2 – slow growth.

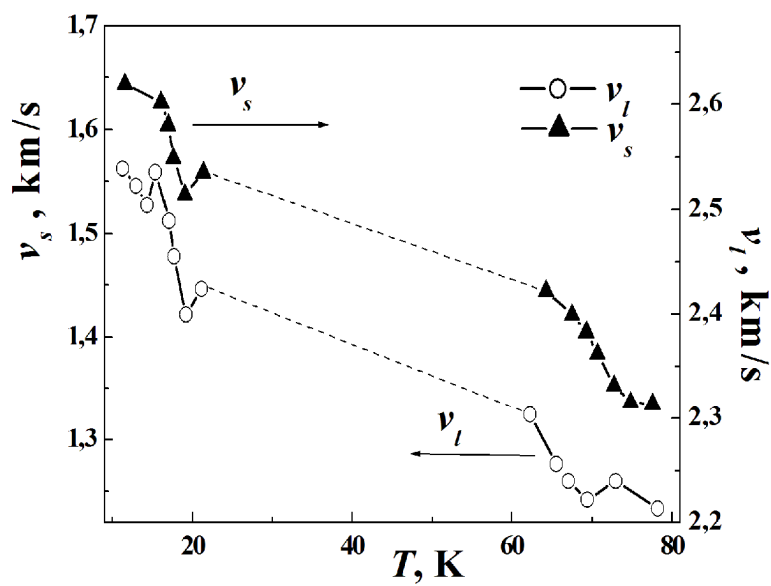

Fig.5. Temperature dependence of velocity of longitudional $v_l$ and transverse $v_s$ sound in solid methane [14] in the vicinity of phase transition temperature and in temperature interval 60-80 K.

It is important to note that although the paper [13] is mentioned in the reviews [3-4], but nothing more, it's results have not been analyzed. As concerns [14], it's results have been cited in the reviews both for velocity of longitudinal and transverse sound. The peculiarities of sound velocity from [14] are clearly seen in corresponding figures of [3-4] like as in Fig.5 of our paper. At that, the data from [14] have been averaged for 4-5 samples, i.e. the ultrasound anomalies are not of random nature. Nevertheless, neither in [14] nor in [3-4] the anomalous behavior of temperature dependence of sound velocity near T=65 K has not been considered.

It should be emphasized that, as distinct from mechanical properties, the anomalies of termodynamical properties of solid methane in analized temperature range are expressed more weakly.

Fig. 6 presents the temperature dependencies of molar volume $V_m$ of solid methane reported in [15-17] by several independent research groups. It is clearly seen that the dependence $V_m(T)$ from [15] shows a peculiarity near 60 K, and the jointing of data from [16] and [17] gives practically a repetition the results of [15]. Anomalies of $V_m(T)$ near 60 K slightly exceed the experimental error of [15-17]. Even though they are small but also can be associated with the transition of **CH$_4$** molecules to a new rotational state.

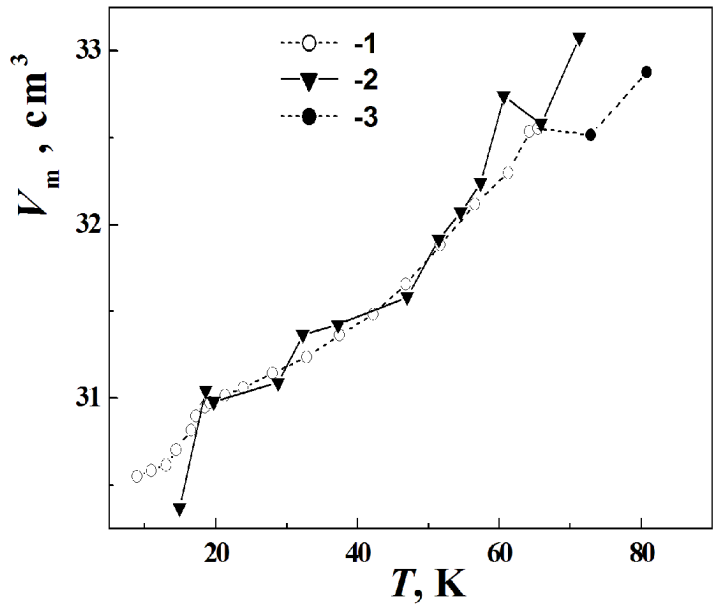

Fig. 6. Temperature dependencies of molar volume for solid methane
according to [15-17]:
1 – [16], 2 – [15], 3 – [17].

Fig. 7 shows the temperature dependencies of volume expansion coefficient $\beta(T)$ according to [16,18-20]. It is seen that divergence of the data of various papers exceeds the possible experimantal errors. Demating of results from [16] and [18] is similar to above $V_m(T)$ behavior. Quite possible that this demating is also related to the transition of methane molecules to a new rotational statIt is interesting to note that although the X-ray data of [19] do not show any visible anomaly on β(T) dependence, the authors of this paper have observed a certain rise of $\beta(T)$ at temperatures above 70 K in comparison with the estimated dependence for ideal lattice (see Fig.3 in [19]). Besides, it is seen in Fig.7 that at temperatures above 70 K values of $\beta(T)$ obtained from dilatometric measurements [20] rises faster in comparison with $\beta(T)$ extracted from X-ray data of [19]. Difference between the dependencies reaches 15% at temperature 89 K (see Table 8.10 from [3]). The authors of [19] explain it by arising of point defects in the sample just as it was observed at premelting temperatures in solidified inert gases (i.e., **Ar, Kr**) so they assert that "solid methane at high temperatures is the closest to crystalline inert gases"[19]. The authors have emphasized that "in this temperature range (T > 70 K) a symmetrical distinction between atoms of inert gases and methane molecules is absent since the last make a practically free rotation in the lattice" [19].

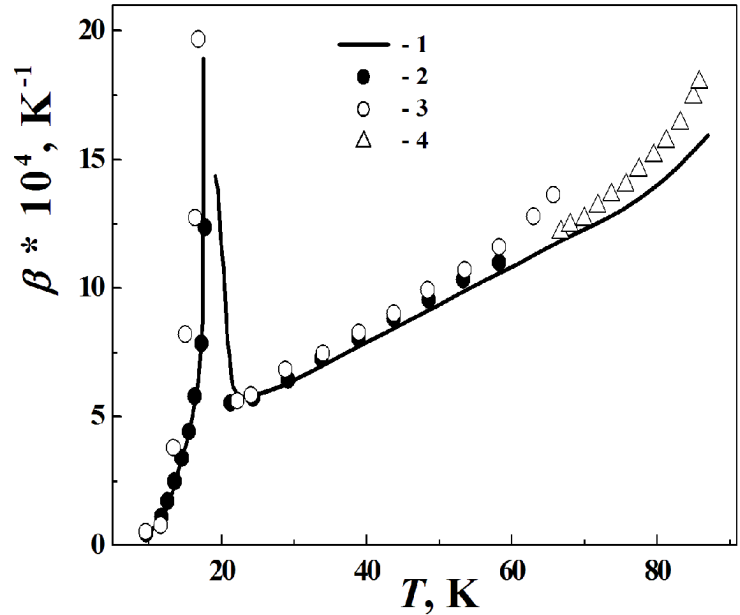

Fig. 7. Temperature dependencies of volume coefficient of thermal expansion $\beta$ for solid methane according to the data of several papers:
1 – [19]; 2 – [20]; 3 – [16]; 4 – [18].

Some more complex situation arises when analyzing the temperature dependence of heat capacity $C_p(T)$. Note that in the analyzed temperature interval ($0.5\ T_{tr}$ – $T_{tr}$), the data of three papers [21-23] are only available. Two earlier of them [21,22] had not reported any noticeable peculiarities in $C_p(T)$ behavior. But the latest paper [23] reports the temperature dependencies of heat capacity for both **CH₄** and **CD₄** and reveals a certain mismatch between these dependencies (see Fig.8).

It has to be noted that, in contrary to **CH₄**, **CD₄** can exist in three phases under equilibrium vapor pressure [3]. High temperature phase **I** and the middle one **II** of **CD₄** are quite similar to the phases **I** and **II** of solid methane. Transition to low temperature phase **III** of **CD₄** is characterized by more complex ordering of **D** atoms relatively to center of mass of carbon atoms accompanied by lowering of symmetry (space group *P-4m2* or *P4₂/mbc* [3]). Therefore, **CD₄** molecules in phase **I**, like to methane ones, make a hindered rotation around the center of mass of carbon atoms. As it can be seen from Fig.8, $C_p(T)$ data cover a whole temperature range of phase **I** and do not show any features.

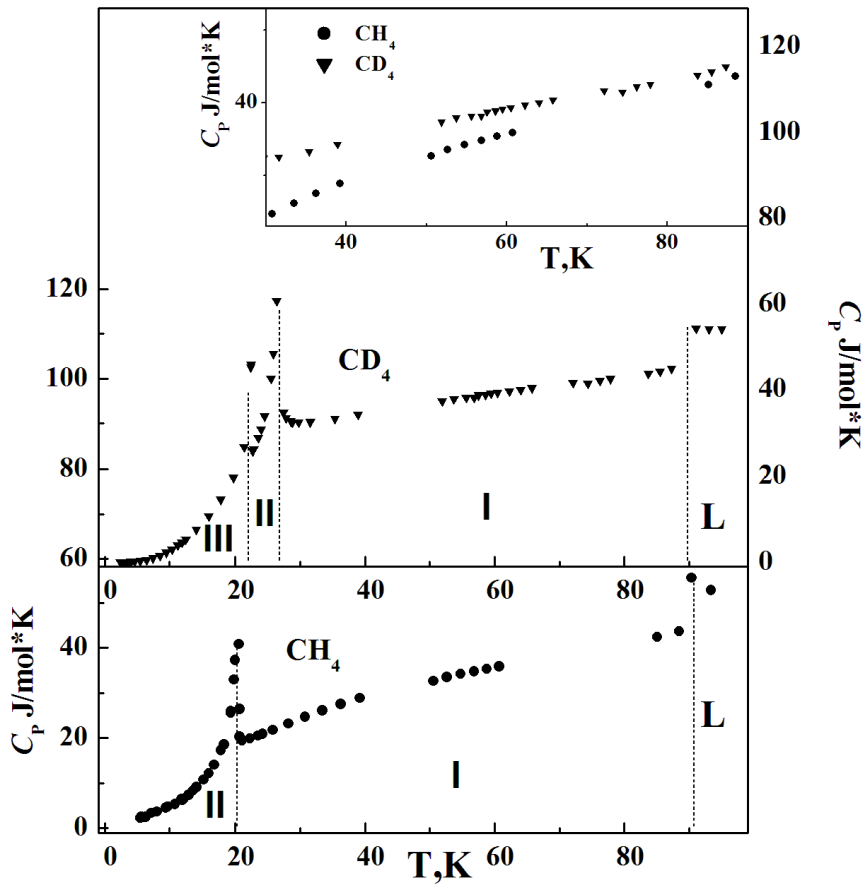

Fig. 8. Temperature dependencies of heat capacity $C_p(T)$ for **CH₄** and **CD₄** according to [23].

Inset in Fig.8 shows the $C_p(T)$ dependencies for both methane and deuteromethane at high temperatures. It can be seen that the data for **CD$_4$** are relatively continious over the all temperature range, and the data for **CH$_4$** are absent between 61 and 85 K. Besides, this lack is not commented in [23]. Probably, the authors have observed a certain anomalies of $C_p(T)$ at these temperatures but, in view of any explanation, exclude these data from the publication.

Therefore, the data on **CH$_4$** heat capacity are not in contradiction with the above assumption about the cooperative behavior of rotational degrees of freedom of methane molecules [12]. The observed break-down of the data is possible related to the transition of quasi-particles of cooperative excitations from quantum (tunnel) to classical (thermoactivation) behavior.

As concerns to **CD$_4$**, it should be remembered that moment of inertia of a **CD$_4$** molecule is twice as much as for a methane molecule. This leads to change of temperature interval of the transition. Because of the lack of a micriscopical theory allowing to establish a relation between the moment of inertia of a single molecule and effective moments of inertia of the cooperative excitations (topons), this temperature interval can not be determined theoretically.

Therefore, a special temperature interval 60-70 K exists in high-temperature range of solid methane. It characterized by anomalies of thermodynamical, mechanical, spectral and elastic properties of crystalline methane that are especially substantial for mechanical and spectral properties. The nature of these anomalies is not studied so far.

Since the rotational degrees of freedom are activated in this temperature interval, it is reasonable to consider its contribution to heat capacity of solid methane and to compare it with experimental data.

## 5. Rotational temperature of solid methane.

Let us suppose that rotational degrees of freedom of crystalline methane are characterized by a sole parameter – rotational temperature $T_r$ that is related to the inertia moment of elementary excitation I by a following expression [24, 25]:

$$T_r = \frac{\hbar^2}{2I}. \tag{1}$$

Further let us call the elementary excitations of rotational degrees of freedom as topons. A topon value I is connected with the inertia moment of methane molecule and intermolecular potentials, but it's real value depends also on thermodynamic conditions of the studied crystalline methane. It is commonly known that the characteristic values of all elementary excitations in solids (for instance, effective electron masses in semiconductors) are substantially dependent on environmental conditions.

Statistical sum of a topon system is given by

$$Z = \sum_{n=0}^{\infty} g_n e^{-\left(\frac{T_r}{T}\right) n(n+1)}, \qquad (2)$$

where $T$ is temperature, $g_n = (2n+1)^2$ is degeneracy multiplicity of the topon's nth state accounting that direction of the rotation axis is not fixed.

Series (2) converges for all temperatures, and for $T < 1.5 T_r$ it is enough to retain as much as ten terms of the series for computing with an accuracy of $10^{-2}\left(T/T_r\right)$. At $T > 1.5 T_r$ the series can be substituted by integral that can be accounted analytically [24, 25].

Note that at $T > 1.5 T_r$ the accounting result obtained by substitution of infinitive series (2) on a sum of the first ten terms are identical to ones for the substitution of the series by an integral. Therefore, in order to determine the temperature dependence of topon's thermal capacity one needs to find the statistical sum (2) with enough number of terms, and then to determine the thermal capacity using the conventional relations of statistical physics. The result of our computing of thermal capacity is shown in Fig.9 in dimensionless variables.

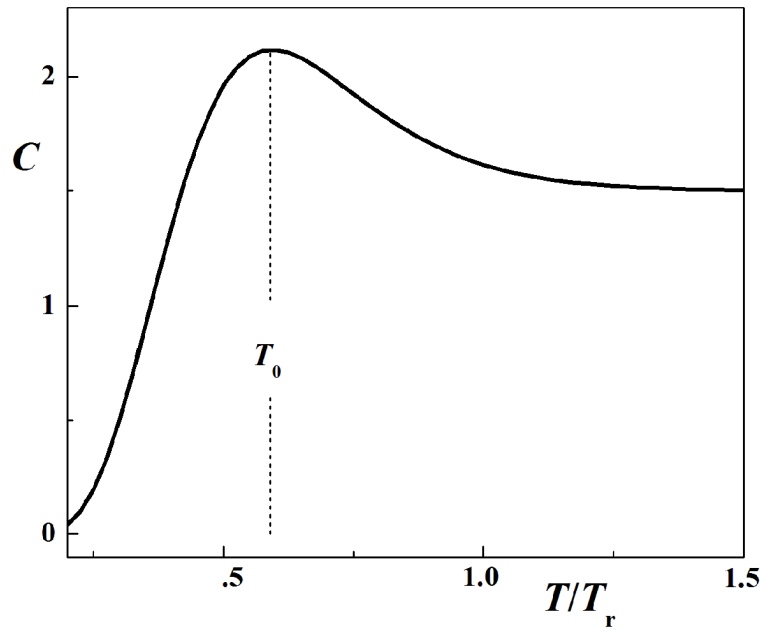

Fig.9. Dependence of specific thermal capacity of topons $C$ on dimensionless temperature $T/T_r$. One hundred terms are involved for computing of the statistical sum (2).

This plot is in qualitative agreement with well-known dependence of thermal capacity on temperature of ideal gas of a top system [24, 25]. A main feature of this plot is existence of thermal capacity maximum at $T_0/T_r \cong 0.59$. Rotational part of thermal capacity at $T > 1.5 T_r$ is well described by the classical theory, and at $T < 1.5 T_r$ the quantum corrections begin to influence. At temperature lowering down to a maximum point $T_0$ these corrections are unessential. But at further temperature decrease the quantum corrections fully suppress the classical thermal capacity. Consequently, the point $T_0$ is practically a boundary between the quantum and classical states of topons in a crystal. Apparently, for the first time an estimation of the boundary $T_0$ between quantum and classical states of a crystal was conducted by Landau and Lifshits [24]. According to their estimation, $T_0$ for methane is approximately 50 K.

It should be noted that according to Tomita's results [3], temperature of the transformation of rotational state of methane molecules is near 60-70 K i.e. has the same order of magnitude as the estimation made in [24]. A corresponding value of the rotational temperature for crystalline methane is $T_r \approx 102 - 119 K$.

There is an independent experimental method for determination of the rotational temperature. It consist in determination of $T_r$ by means of spectroscopic data both under the Earth conditions and at the planets (and their satellites) of the Solar system [26-28]. Rotational temperature of solid methane obtained from the spectroscopic data for Saturn lies between 122 and 142 K [26]. The same measurements for Jupiter [27,28] yield the rotational temperature value between 150 and 230 K. Difference between the values for Saturn and Jupiter, probably, is associated with different physical conditions at these planets.

Therefore, rotation of molecules in solid methane can be represented as a result of collectivization of the rotational degrees of freedom of the molecules.

Theoretical estimation of rotational temperature $T_r$ (the only parameter characterizing the cooperative rotational degrees of freedom of solid methane) is very difficult. However, it can be derived from an analysis of a certain experimental data, in particular, spectroscopic ones. A whole of various independent experimental data on abnormal behavior of solid methane at temperatures 60-70 K can be explained in the frameworks of the conception of topons that are cooperative excitations of rotational degrees of freedom of methane molecules.

## 6. Analysis of contribution of rotational component to thermal capacity of methane

As it has been mentioned above, the data of three papers [21-23] are only available in the analyzed temperature interval $0.5 T_{tr} - T_{tr}$. Since the results of these papers have certain mismatches but at the same time it is possible to use the averaged data of the papers to estimate the contribution of rotational component to a total heat capacity of solid methane.

Note that a problem of separation of the contributions of various kinds of lattice oscillations to heat capacity is a rather complex task. But in case of simplest molecular crystals, a simplified method of separate consideration of contributions from translational and rotational molecular degrees of freedom to thermal capacity is usually applied. With that, the Debye's approximation is most commonly used for account of translational contributions to thermal capacity. A difference between the thermal capacity for constant volume ($C_v$) and accounted one in Debye's approximation is commonly considered as contribution of rotational degrees of freedom to heat capacity. Firstly, the contribution of rotational degrees of freedom to heat capacity for solid methane was obtained in [19] in the frameworks of above conception. Using their data on volume expansion coefficient ($\beta$) amd molar volume ($V_m$), the values of velocities of longitudinal $v_l$ and transverse $v_s$ sound and adiabatic compressibility ($\chi_s$) [20,29], and heat capacity at constant pressure ($C_P$) [21-23], authors of [19] have accounted the heat capacity $C_v$ and have separated the contribution of rotational degrees of freedom $C_{rot}$ from it. However, they used the volume expansion coefficients obtained from X-ray data [25] that at T> 65 K are far less than $\beta$ values obtained from dilatometric studies [18,20]. At that, the difference between them quickly raises with temperature growth (see Fig.10).

Taking into account that

$$C_P - C_V = \beta^2 V_m T / \chi \qquad (3)$$

then $C_{rot}$ obtained in [19] at T> 65 K are obviously overstated (see Fig.10).

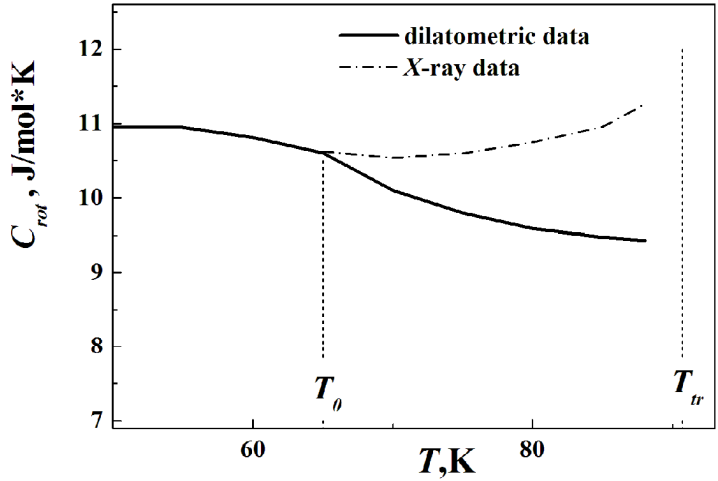

Fig.10. Contribution of rotational degrees of freedom to thermal capacity of solid methane at temperatures above 50 K.

As it clearly seen in Fig.10, temperature dependence of $C_{rot}$ is in qualitative agreement with the theoretical one in Fig.9 at temperatures above 60 K.

## 7. Conclusions.

The analysis of available experimental data about the mechanical, structural, thermophysical and elastic properties of solid methane in temperature range above $0.5 \cdot T_{tr}$ under equilibrium vapor pressure shows that the most of considered papers reports the anomalies of the studied properties at temperatures 60-70 K or have not the reliable data at this temperature range.

On the ground of analysis of numerous experimental data and theoretical investigation of rotational degrees of freedom behavior in high-temperature phase of solid methane, a consception of topons (cooperative excitations of rotational degrees of freedom in solid methane) has been proposed. We have shown that the anomalies of studied properties reported in numerous papers for temperature interval 60-70 K, can be successfully explained in the frameworks of this approach.

Taking into account the results of Tomita [5] where he firstly experimentally observed a sharp rise of activation energy of molecular rotation (~ 20 times) in solid methane at temperatures 60-70 K, we can conclude that the observed anomalies are caused by a transition of methane molecules to a new rotational state. This state is associated with a transition of cooperative degrees of freedom in solid methane from quantum behavior to classic one [12] at these temperatures.

Such a sudden change of rotational state of methane molecules considerably affects on the **NMR** and **LFIF** spectra as well as on mechanical characteristics of solid methane. More weakly this phenomenon is displayed in temperature dependencies of thermodynamic parameters of solid methane.

The transformation of rotational state of the methane molecules in temperature interval 60-70 K must considerably weaken the intermolecular forces in solid methane. This, in turn, affects its mechanical characteristics. Therefore, results of the present paper may be of great importance, in particular, when analyzing the astrophysical data about the surface of Solar system planets or their satellites.